\documentclass[acmsmall,screen,authorversion,nonacm]{acmart}

\usepackage{listings}
\usepackage{color}
\definecolor{bluekeywords}{rgb}{0.13, 0.19, 0.7}
\definecolor{greencomments}{rgb}{0.1, 0.5, 0.2}
\definecolor{redstrings}{rgb}{0.8, 0.15, 0.1}
\definecolor{graynumbers}{rgb}{0.5, 0.5, 0.5}
\definecolor{subtlegray}{rgb}{0.98, 0.98, 0.98}
\usepackage{lstautogobble}
\usepackage{tikz}
\usetikzlibrary{quantikz}

\AtBeginDocument{%
  \providecommand\BibTeX{{%
    \normalfont B\kern-0.5em{\scshape i\kern-0.25em b}\kern-0.8em\TeX}}}

\acmConference[PLanQC 2022]{Third International Workshop on Programming Languages for Quantum Computing}{September 11-16, 2022}{Ljubljana, Slovenia}
\acmBooktitle{Third International Workshop on Programming Languages for Quantum Computing}

\begin{document}

\title[{Quantum-Circuit Transformations}]{Transformations for Accelerator-based Quantum Circuit Simulation in Haskell}

\author{Youssef Moawad}
\email{}
\affiliation{%
  \institution{University of Glasgow}
  \streetaddress{School of Computing Science, 18 Lilybank Gardens.}
  \city{Glasgow}
  \state{}
  \country{UK}
  \postcode{G128QQ}
}
\author{Wim Vanderbauwhede}
\email{}
\affiliation{%
  \institution{University of Glasgow}
  \streetaddress{School of Computing Science, 18 Lilybank Gardens.}
  \city{Glasgow}
  \state{}
  \country{UK}
  \postcode{G128QQ}
}
\author{René Steijl}
\email{}
\affiliation{%
  \institution{University of Glasgow}
  \streetaddress{James Watt School of Engineering, James Watt South Building.}
  \city{Glasgow}
  \state{}
  \country{UK}
  \postcode{G128QQ}
}


\begin{abstract}
For efficient hardware-accelerated simulations of quantum circuits, we can define hardware-specific quantum-circuit transformations. We use a functional programming approach to create a quantum-circuit analysis and transformation method implemented in Haskell. This tool forms a key part of our larger quantum-computing simulation toolchain. As an example of hardware acceleration, we discuss FPGA-based simulations of selected quantum arithmetic circuits, including the transformation steps to optimise the hardware utilisation. Future development steps in the Haskell-based analysis and transformation tool are outlined. The described toolchain can be found on GitHub: https://github.com/DevdudeSami/fqt.
%

\end{abstract}





\maketitle

\section{Introduction}

Efficient simulation of quantum computers is essential for the development of quantum algorithms and for further development of quantum hardware. Quantum computers promise to deliver up to exponential complexity improvements for certain algorithms\cite{nielsen_quantum_2010}; however, this is also where the difficulty in simulating them arises. The state of a coherent $n$-qubit quantum register in a quantum processor can be defined using $2^{n}$ complex numbers, also termed the quantum state vector. In contrast to the register in a classical computer, the quantum state in a coherent qubit register is in a state of superposition until quantum measurement operations are performed. Measurement (partially) collapses the quantum state superposition, and the complex amplitudes define the likelihood of finding a given state after measurement. Quantum superposition is the key principle creating the potential speed-up for most quantum algorithms.

The Quantum Circuit Model is the most common model for interacting with current quantum hardware and reasoning about quantum algorithms. In this work, Qubit-Wise Multiplication (QWM) is chosen as the baseline simulation method. While this requires the storage of the full state-vector, $2^{n}$ complex amplitudes, it does give the most control and allows full inspection of the state during computation. To process one quantum gate in QWM, the entire state vector is updated. Data-locality during these operations is not guaranteed, as pairs of amplitudes have to be accessed with strides that grow exponentially depending on the target qubit. However, for each gate, due to the implied quantum parallelism, these operations can be executed in parallel. 

Classical simulations of quantum circuits are most commonly performed on multicore computers and clusters\cite{de_raedt_massively_2007, de_avila_state---art_2019}. A range of high-performance simulators exists. 
Typically, the quantum-circuit implementation of quantum algorithms is represented by a Domain Specific Language (e.g. QASM).
Quantum computer simulations have also been performed on GPUs\cite{gutierrez_quantum_2010, kelly_simulating_2018} and FPGAs\cite{khalid_fpga_2004, aminian_fpga-based_2008, lee_fpga-based_2016, pilch_fpga-based_2019}. The underlying idea of such {\it hardware-accelerated} simulations is to use the specific hardware features to process the quantum circuit more efficiently. In our ongoing research, we focus on FPGAs (Field Programmable Gate Arrays) as accelerators. FPGAs are programmable circuits that allow to construct highly parallel architectures that closely mimic the properties of quantum computing. To allow us to investigate different architectures, we developed a quantum simulation toolchain that includes a Quantum Circuit Analysis and Transformation tool. We target quantum algorithms for computational science and engineering applications. In contrast to a significant body of work focusing on general and even randomized circuits, the present work aims to explicitly take advantage of the quantum-circuit structure using the knowledge of the domain experts developing the algorithms.
Specifically, our work differs in scope and context from previous quantum computing tool chains employing Functional Programming, e.g. Quipper and Microsoft's  Q\# and Liquid.

The key contributions of this paper can be summarized as follows:
\begin{itemize}
    \item Discussion of the design and implementation of our toolchain. 
    \item An implementation of an FPGA-based quantum circuit simulator embedded in our toolchain and its embedded DSL 
    \item A new circuit optimisation technique based on the reduction of input and workspace qubits for generating specialised circuits with fewer qubits, reducing the memory space required for the full state-vector simulation approach;
\end{itemize}
%

%


%


%

\section{Simulation of Quantum Circuits}

\subsection{CPU/GPU Simulation}

Extensive research has resulted in a range of highly-optimized quantum computer simulators for multi-core and distributed computing architectures. Intel Quantum Simulator (IQS, formerly known as qHiPSTER) \cite{smelyanskiy_qhipster_2016} is a quantum simulator optimised for multi-node systems. ProjectQ \cite{steiger_projectq_2018} offers a modular compiler engine that can optimise at different levels of abstraction defined by the user, and includes optimised tools for local simulation of circuits. JUMPIQCS \cite{de_raedt_massively_2007}  is a Fortran 90-based simulator that utilises MPI for distribution.  Qrack \cite{strano_qrack_nodate} and QCGPU \cite{kelly_simulating_2018} are cross platform OpenCL-based full state-vector simulators.



\subsection{FPGA Simulation}

The simulation of quantum computers and, specifically, quantum-circuit implementations of algorithms on FPGAs has been the topic of more recent research works. Examples of works focusing on FPGAs include\cite{khalid_fpga_2004}, \cite{aminian_fpga-based_2008}, \cite{conceicao_efficient_2015}, \cite{lee_fpga-based_2016}, \cite{mahmud_scalable_2018}, \cite{pilch_fpga-based_2019}, \cite{mahmud_efficient_2020}, \cite{khalid_fpga_2021}, and \cite{bonny_emulation_2020}.

\subsection{Simulator architecture for FPGAs}




Our main goals for developing an FPGA simulator for quantum circuits are: \emph{universality} (ability to simulate any theoretical gate), \emph{reuseability} (a recompilation process should not be necessary between different circuit runs), and \emph{scalability} (we should be able to simulate any feasible number of qubits without recompiling). We achieve universality by making sure the system has built-in at least a universal set of quantum gates. Our current architecture (implemented in OpenCL and tested with Intel's AOCL compiler) stores the state vector in FPGA DRAM and compute kernels corresponding to each quantum gate access the memory to perform the necessary computations. Since in general each gate application needs to access the entire memory space, we perform gate applications sequentially and attempt to optimise the performance of the application of a general gate.

\section{Haskell Toolchain}

Debugging complex quantum circuits at the level of our FPGA instruction set can be very tedious and so several higher level languages exist for expressing quantum algorithms, including Quipper \cite{green_introduction_2013}, OpenQASM 3 \cite{cross_openqasm_2021}, and Microsoft Q\# \cite{hooyberghs_q_2022}. We decided to include a custom eDSL with our toolchain to maintain control and facilitate future development of architecture-specific optimisations in the instruction set. However, implementing frontends for these already existing high-level languages would allow for a tighter integration with the current ecosystem.

The main contribution of this work is the introduction of a Haskell-based toolchain and eDSL for specifying and compiling quantum circuits for an FPGA-based architecture.

\subsection{Core}

The Core modules of the toolchain provide the constructs used in the specification of a quantum circuit. This includes primary definitions for types used throughout the tool and an inner Circuit type to represent a circuit over an indexed quantum register (i.e. very close to what the FPGA will actually process). On top of this Circuit type, the eDSL constructs are defined. This includes utilities for referring to qubits by names instead of indices (essentially defining arbitrary "quantum pointers"), arbitrary controls and negative controls defined over a gate or a set of gates, circuit chaining, looping and tiling subcircuits, etc. 

\begin{lstlisting}[language=Haskell, caption=Generic input size full Cuccaro adder and square circuit example implementation in the presented Haskell eDSL., label=code_nBitSquareQC]
fullAdd :: QReg -> QReg -> Qu -> Qu -> Circ 
  fullAdd in1 in2 c z = if length in1 /= length in2 then error "fullAdd: Input qubit register lengths must be identical." else let
    combinedRegister = c : interleave in2 in1
    in 
      ladderQC 2 3 maj combinedRegister ++
      cnot (last in1) z ++
      reverseLadderQC 2 3 unmaj combinedRegister

\end{lstlisting}

The Core modules also include two simulators, implemented in Haskell, for convenience. One is a general full state-vector QWM simulator with no optimisations which was used to model early QWM-based simulators. Additionally, a "logic" simulator is also provided which can simulate circuits containing only the $NOT$ gate with any number of controls; this is useful for quickly debugging circuits which operate only in the computational basis. Since only one state is set at any point in the circuit, simulating such circuits can be performed in linear time and memory.

\subsection{Testing}

As is the case with classical software, effective unit testing of quantum circuits is very important. This involves running quantum circuits with different input states and checking that the output states fit some expectations. To facilitate automating this task, a testing framework is provided with the toolchain. Currently, the side of this which interacts with the FPGA simulator takes qubit preparations which it uses to compute the input state of the circuit. The circuit is then simulated by the FPGA and the output state is passed back into the toolchain (currently this is manual but automation is planned), decoded, and checked against the test expectations. There is also an option to run the tests against the included logic simulator, which proved particularly useful for debugging complex computational-basis circuits.


\subsection{Compiler}

The compiler modules include functions and tools for going from the eDSL representation to the FPGA instructions. Alternatively, the compiler can also read a QASM-like file specifying the circuit.

\begin{figure}[htbp]
  \centering
  \includegraphics[width=1\columnwidth]{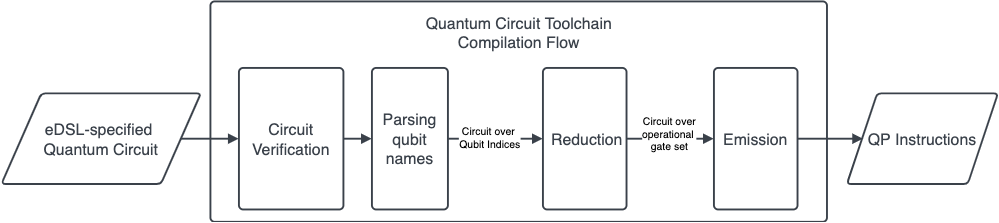}
  \caption{eDSL to FPGA input representation.}
  \label{fig_eDSL2QP}
\end{figure}

The compilation process is demonstrated in Figure \ref{fig_eDSL2QP}. First, the specified circuit is verified, ensuring all qubits used are valid (have an index in the register) and no gates are specified with invalid target/controls. Then the named qubit identifiers are parsed away and the qubits are mapped to an index in the quantum register. At this point in the process, some constructs are still available to the tool which would not necessarily be available to the FPGA (like direct calls to a SWAP gate, or a high number of controls), which need to be reduced away. SWAP gates are replaced with their equivalent CNOT specifications, negative controls are reduced by negating the control qubits before and after the gate, and gates with a higher number of controls than is supported are expanded to several gates with fewer controls. This results in a circuit which is ready to be converted to a QP (Quantum Problem) file which is simply a list of integers specifying the circuit. Taking into account the maximum number of controls allowed by the architecture, each emitted gate consists of its opcode, target qubit, followed by a constant number of controls. The resulting list is then written to disk, ready to be read by the simulator host.

\subsection{Circuit Qubit Reduction}

Quantum circuits representing quantum algorithms which employ computational-basis encoding of can have some qubits reduced out by generating two different circuits for each possible qubit input. In this way, the total memory required for one run of the circuit is reduced by half for each qubit reduced out. Qubits which are only used as controls throughout the circuit are ideal candidates for this type of reduction, and the toolchain provides functionality to automate this. While this approach is useful for reducing the total memory required across any platform, it is especially good for an FPGA which would, in theory, be able to run both (for one reduced qubit) circuits concurrently on completely independent memory spaces, which improves data locality.

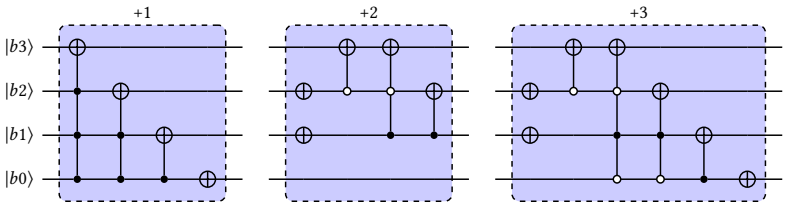
\begin{figure}[h]
    \centering
\begin{tikzpicture}
\node[scale=0.7]{
\begin{quantikz}
\lstick{$\ket{b3}$} &\targ{}\gategroup[4,steps=4,style={dashed, rounded corners,fill=blue!20, inner xsep=2pt},background]{{\sc +1}} &\qw &\qw &\qw &\qw & &\qw\gategroup[4,steps=4,style={dashed, rounded corners,fill=blue!20, inner xsep=2pt},background]{{\sc +2}} &\targ{} &\targ{} &\qw &\qw & &\qw\gategroup[4,steps=6,style={dashed, rounded corners,fill=blue!20, inner xsep=2pt},background]{{\sc +3}} &\targ{} &\targ{} &\qw &\qw &\qw &\qw\\
\lstick{$\ket{b2}$} &\ctrl{-1} &\targ{} &\qw &\qw &\qw & &\targ{} &\octrl{-1} &\octrl{-1} &\targ{} &\qw & &\targ{} &\octrl{-1} &\octrl{-1} &\targ{} &\qw &\qw &\qw \\
\lstick{$\ket{b1}$} &\ctrl{-1} &\ctrl{-1} &\targ{} &\qw &\qw & &\targ{} &\qw &\ctrl{-1} &\ctrl{-1} &\qw & &\targ{} &\qw &\ctrl{-1} &\ctrl{-1} &\targ{} &\qw &\qw \\
\lstick{$\ket{b0}$} &\ctrl{-1} &\ctrl{-1} &\ctrl{-1} &\targ{} &\qw & &\qw &\qw &\qw &\qw &\qw & &\qw &\qw &\octrl{-1} &\octrl{-1} &\ctrl{-1} &\targ{} &\qw
\end{quantikz}
};
\end{tikzpicture}
\caption{Cuccaro 4-qubit modulo adder: example kernels for '+1', '+2' and '+3'}
\label{fig_DivCuccaro_MOD4_kernels}
\end{figure}

%
As an example quantum-circuit transform, the Cuccaro modulo $4$-qubit adder is considered. The original $9$-qubit circuit is shown in Figure \ref{fig_DivCuccaro_MOD4}, where $\ket{a3|a2|a1|a0}$ and $\ket{b3|b2|b1|b0}$ represent the two $4$-qubit inputs, so that $\ket{a3|a2|a1|a0}$ remains unchanged and after completion $\ket{b3|b2|b1|b0}$ holds the summation (for clarity renamed as $\ket{s3|s2|s1|s0}$). Since the qubits in $\ket{a3|a2|a1|a0}$ are the same after completion, a transformation can be defined so that circuits for specific inputs $\ket{a3|a2|a1|a0}$ are created. Also, if we move the input qubits that remain the same to the top of the circuit (here the most significant qubits), then $4$-qubit circuits can be created that add a specific integer in $4$-qbit binary representation. This transformation is illustrated here, where the qubits $\ket{a3|a2|a1|a0}$ are first moved to top of circuit, as shown in Figure \ref{fig_DivCuccaro_MOD4}. Then, by specializing $\ket{a3|a2|a1|a0}=\ket{0001}$, $\ket{a3|a2|a1|a0}=\ket{0010}$ or $\ket{a3|a2|a1|a0}=\ket{0011}$, the three example $4$-qubit reduced circuits shown in Figure \ref{fig_DivCuccaro_MOD4_kernels} can be created. One of the aims of the Haskell-based transformation tool is to perform this type of transformation automatically for more complex circuits where 'constant' qubits acting (predominantly) as control to gate operations on other qubits can be identified.

\begin{figure}[h]
    \centering
\begin{tikzpicture}
\node[scale=0.4]{
\begin{quantikz}
\lstick{$\ket{a3}$} &\qw &\qw &\qw &\qw &\qw &\qw &\qw &\qw &\qw &\ctrl{1} &\qw &\qw &\qw &\qw &\qw &\qw &\qw &\qw &\qw &\qw &\qw &\rstick{$\ket{a3}$} \\
\lstick{$\ket{b3}$} &\qw &\qw &\qw &\qw &\qw &\qw &\qw &\qw &\qw &\targ{} &\targ{} &\qw &\qw &\qw &\qw &\qw &\qw &\qw &\qw &\qw &\qw &\rstick{$\ket{s3}$} \\
\lstick{$\ket{a2}$} &\qw &\qw &\qw &\qw &\qw &\qw &\ctrl{1} &\ctrl{2} &\targ{} &\qw &\ctrl{-1} &\targ{} &\ctrl{2} &\qw &\qw &\qw &\qw &\qw &\qw &\qw &\qw &\rstick{$\ket{a2}$} \\
\lstick{$\ket{b2}$} &\qw &\qw &\qw &\qw &\qw &\qw &\targ{} &\qw &\ctrl{-1} &\qw &\qw &\ctrl{-1} &\qw &\targ{} &\qw &\qw &\qw &\qw &\qw &\qw &\qw &\rstick{$\ket{s2}$} \\
\lstick{$\ket{a1}$} &\qw &\qw &\qw &\ctrl{1} &\ctrl{2} &\targ{} &\qw &\targ{} &\ctrl{-1} &\qw &\qw &\ctrl{-1} &\targ{} &\ctrl{-1} &\targ{} &\ctrl{2} &\qw &\qw &\qw &\qw &\qw &\rstick{$\ket{a1}$} \\
\lstick{$\ket{b1}$} &\qw &\qw &\qw &\targ{} &\qw &\ctrl{-1} &\qw &\qw &\qw &\qw &\qw &\qw &\qw &\qw &\ctrl{-1} &\qw &\targ{} &\qw &\qw &\qw &\qw &\rstick{$\ket{s1}$} \\
\lstick{$\ket{a0}$} &\ctrl{1} &\ctrl{2} &\targ{} &\qw &\targ{} &\ctrl{-1} &\qw &\qw &\qw &\qw &\qw &\qw &\qw &\qw &\ctrl{-1} &\targ{} &\ctrl{-1} &\targ{} &\ctrl{2} &\qw &\qw &\rstick{$\ket{a0}$}\\
\lstick{$\ket{b0}$} &\targ{} &\qw &\ctrl{-1} &\qw &\qw &\qw &\qw &\qw &\qw &\qw &\qw &\qw &\qw &\qw &\qw &\qw &\qw &\ctrl{-1} &\qw &\targ{} &\qw &\rstick{$\ket{s0}$} \\
\lstick{$\ket{c}=\ket{0}$} &\qw &\targ{} &\ctrl{-1} &\qw &\qw &\qw &\qw &\qw &\qw &\qw &\qw &\qw &\qw &\qw &\qw &\qw &\qw &\ctrl{-1} &\targ{} &\ctrl{-1} &\qw &\rstick{$\ket{c}=\ket{0}$}
\end{quantikz}
};
\end{tikzpicture}\\
\mbox{$4$-qubit modulo adder: original circuit}\\
\begin{tikzpicture}
\node[scale=0.4]{
\begin{quantikz}
\lstick{$\ket{a3}$} &\qw &\qw &\qw &\qw &\qw &\qw &\qw &\qw &\qw &\ctrl{4} &\qw &\qw &\qw &\qw &\qw &\qw &\qw &\qw &\qw &\qw &\qw &\rstick{$\ket{a3}$} \\
\lstick{$\ket{a2}$} &\qw &\qw &\qw &\qw &\qw &\qw &\ctrl{4} &\ctrl{1} &\targ{} &\qw &\ctrl{3} &\targ{} &\ctrl{1} &\qw &\qw &\qw &\qw &\qw &\qw &\qw &\qw &\rstick{$\ket{a2}$} \\
\lstick{$\ket{a1}$} &\qw &\qw &\qw &\ctrl{4} &\ctrl{1} &\targ{} &\qw &\targ{} &\ctrl{-1} &\qw &\qw &\ctrl{-1} &\targ{} &\ctrl{3} &\targ{} &\ctrl{1} &\qw &\qw &\qw &\qw &\qw &\rstick{$\ket{a1}$} \\
\lstick{$\ket{a0}$} &\ctrl{4} &\ctrl{5} &\targ{} &\qw &\targ{} &\ctrl{-1} &\qw &\qw &\qw &\qw &\qw &\qw &\qw &\qw &\ctrl{-1} &\targ{} &\ctrl{3} &\targ{} &\ctrl{5} &\qw &\qw &\rstick{$\ket{a0}$} \\
\lstick{$\ket{b3}$} &\qw &\qw &\qw &\qw &\qw &\qw &\qw &\qw &\qw &\targ{} &\targ{} &\qw &\qw &\qw &\qw &\qw &\qw &\qw &\qw &\qw &\qw &\rstick{$\ket{s3}$} \\
\lstick{$\ket{b2}$} &\qw &\qw &\qw &\qw &\qw &\qw &\targ{} &\qw &\ctrl{-3} &\qw &\qw &\ctrl{-3} &\qw &\targ{} &\qw &\qw &\qw &\qw &\qw &\qw &\qw &\rstick{$\ket{s2}$} \\
\lstick{$\ket{b1}$} &\qw &\qw &\qw &\targ{} &\qw &\ctrl{-3} &\qw &\qw &\qw &\qw &\qw &\qw &\qw &\qw &\ctrl{-3} &\qw &\targ{} &\qw &\qw &\qw &\qw &\rstick{$\ket{s1}$} \\
\lstick{$\ket{b0}$} &\targ{} &\qw &\ctrl{-4} &\qw &\qw &\qw &\qw &\qw &\qw &\qw &\qw &\qw &\qw &\qw &\qw &\qw &\qw &\ctrl{-4} &\qw &\targ{} &\qw &\rstick{$\ket{s0}$} \\
\lstick{$\ket{c}=\ket{0}$} &\qw &\targ{} &\ctrl{-1} &\qw &\qw &\qw &\qw &\qw &\qw &\qw &\qw &\qw &\qw &\qw &\qw &\qw &\qw &\ctrl{-1} &\targ{} &\ctrl{-1} &\qw &\rstick{$\ket{c}=\ket{0}$}
\end{quantikz}
};
\end{tikzpicture}\\
\mbox{$4$-qubit modulo adder: re-arranged circuit with $\ket{a3|a2|a1|a0}$ as most-significant qubits}
\caption{Cuccaro 4-qubit modulo adder: original circuit and re-arranged circuit}
\label{fig_DivCuccaro_MOD4}
\end{figure}

\section{Conclusion and Future Work}

We presented a Haskell-based toolchain for compilation and optimisation of quantum circuits for the purpose of FPGA-based simulation. Currently, our toolchain facilitates encoding, debugging, and unit testing circuits before compiling to a QASM format specific to our FPGA architectures. Preliminary implementations of the described circuit qubit reduction optimisation are also included. We demonstrate how these circuit reduction techniques can be applied to computational-basis circuits to facilitate splitting them to smaller circuits; reducing the memory required for simulation and benefiting customised parallel accelerators. Future work will focus on circuit analysis methods to facilitate implementing FPGA-specific optimisations including dynamic reconfiguration of the FPGA bitstream at runtime, cost-modelling of the memory and caches, and further automation of Circuit-Qubit Reduction transformations.

\bibliographystyle{plain}
\bibliography{references.bib}

\end{document}